\documentclass[12pt,a4paper,english,nofootinbib]{revtex4}
\usepackage{lmodern}
\usepackage{lmodern}

\usepackage[T1]{fontenc}
\usepackage[latin9]{inputenc}
\usepackage{babel}
\usepackage{mathrsfs}
\usepackage{amsmath}
\usepackage{amssymb}
\usepackage{graphicx}
\usepackage{esint}
\usepackage[unicode=true,pdfusetitle,
 bookmarks=true,bookmarksnumbered=false,bookmarksopen=false,
 breaklinks=false,pdfborder={0 0 1},backref=false,colorlinks=false]
 {hyperref}

\makeatletter

\newcommand{\lyxmathsym}[1]{\ifmmode\begingroup\def\b@ld{bold}
  \text{\ifx\math@version\b@ld\bfseries\fi#1}\endgroup\else#1\fi}

\@ifundefined{textcolor}{}
{%
 \definecolor{BLACK}{gray}{0}
 \definecolor{WHITE}{gray}{1}
 \definecolor{RED}{rgb}{1,0,0}
 \definecolor{GREEN}{rgb}{0,1,0}
 \definecolor{BLUE}{rgb}{0,0,1}
 \definecolor{CYAN}{cmyk}{1,0,0,0}
 \definecolor{MAGENTA}{cmyk}{0,1,0,0}
 \definecolor{YELLOW}{cmyk}{0,0,1,0}
}

\usepackage{babel}
\usepackage{babel}
\usepackage{babel}
\usepackage{babel}
\usepackage{babel}
\usepackage{babel}
\usepackage{babel}
\usepackage{babel}
\usepackage{babel}
\usepackage{babel}
\usepackage{babel}
\usepackage{babel}
\usepackage{babel}
\usepackage{babel}
\usepackage{babel}
\usepackage{babel}
\usepackage{babel}
\usepackage{babel}
\usepackage{babel}

\usepackage{babel}
\def\b{\begin{equation}}
	\def\e{\end{equation}}

\@ifundefined{textcolor}{}{%
	\definecolor{BLACK}{gray}{0}
	\definecolor{WHITE}{gray}{1}
	\definecolor{RED}{rgb}{1,0,0}
	\definecolor{GREEN}{rgb}{0,1,0}
	\definecolor{BLUE}{rgb}{0,0,1}
	\definecolor{CYAN}{cmyk}{1,0,0,0}
	\definecolor{MAGENTA}{cmyk}{0,1,0,0}
	\definecolor{YELLOW}{cmyk}{0,0,1,0}
}

\usepackage{latexsym}\usepackage{bm}

\makeatother

\begin{document}
\title{Basics of Apparent Horizons in Black Hole Physics}
\author{{\normalsize{}{}{}{}{}{}{}{}{}{}{}{}{}{}{}Emel Altas}}
\email{emelaltas@kmu.edu.tr}

\affiliation{Department of Physics,\\
 Karamanoglu Mehmetbey University, 70100, Karaman, Turkey}
\author{{\normalsize{}{}{}{}{}{}{}{}{}{}{}{}{}{}{}Bayram Tekin}}
\email{btekin@metu.edu.tr}

\affiliation{Department of Physics,\\
 Middle East Technical University, 06800, Ankara, Turkey}
\date{{\normalsize{}{}{}{}{}{}{}{}{{}{}{}{}\today}}}

\maketitle
\footnote{This paper is written for Prof. Tekin Dereli's 70th Birthday Festschrift.}Event
Horizon, a null hypersurface defining the boundary of the black hole
region of a spacetime, is not particularly useful for evolving black
holes since it is non-local in time. Instead, one uses the more tangible
concept of Apparent Horizon for dynamical black holes out there in
the sky that do all sorts of things: evolve, merge and feed on the
environment. Event Horizon, being a gauge-independent, global property
of the total spacetime is easy to define and locate in the stationary
case; on the other hand, Apparent Horizon depends on the embedding of the surface in spacetime 
and hence it is somewhat tricky to define. But for numerical simulations in
General Relativity, locating the Apparent Horizon helps one to excise
the black hole region and the singularity to have a stable computation. Moreover, for stationary solutions the two horizons match.
Here we give a detailed pedagogical exposition of the subject and
work out the non-trivial case of a slowly moving and spinning black
hole.

\section{Introduction}

Stationary (Kerr) and static (Schwarzschild) black hole solutions
of General Relativity have rather dull lives: stationary ones do the
same thing, static ones do nothing as observed by an observer outside
the black hole. While these vacuum solutions obtained in an isolated
universe serve as our starting point for a more physical and detailed
understanding of actual astrophysical black holes, the latter are
almost never isolated: the black holes out in the sky have accretion
disks, companion stars, neutron stars or black holes. Black holes
feed on their environment and grow; in fact they are the most dynamical
parts of the vacuum. As the first LIGO/VIRGO gravitational wave detection
showed \cite{merger}, black holes can grow feeding on other black
holes: cannibalistic behavior of these objects-highly curved vacua-could
explain the existence of intermediate mass black holes.

As the astrophysical black holes evolve, concepts such as the Event
Horizon defined easily for eternal black holes are not clearly adequate for
us, the transient observers. Recall that the Event Horizon (${\mathcal{H}}$)
of a stationary black hole is a co-dimension one null hypersurface
in the totality of the spacetime defined as the \textit{boundary of
the black hole region from which time-like or light-like geodesics
cannot reach future null infinity }\cite{Eric0}. Stated in a different
way: it is the boundary of the region which is not in the causal past
of the future null-infinity. This says that the Event Horizon
is a global property of the totality of events which is all of the spacetime.
Therefore, one cannot locate the Event Horizon with local experiments in a finite
interval of time. In this respect, it is apt to say that the Event
Horizon Telescope detected the environment of the black hole from
which one can see at best the cross-section of the Event Horizon, not the
Event Horizon itself.

For dynamical black holes one invents the more useful concept of
the `{}`Apparent Horizon'' \cite{Hawking-Ellis}, a co-dimension
two spatial surface (hence local in time), which, unfortunately, in
general does not carry geometric invariant data as the Event Horizon
but it contains sufficient information regarding the possible formation
of an Event Horizon in the future that it pays to describe it in detail. In numerical
relativity computations, detection of a black hole region is best done with Apparent
Horizons. Within the context of General Relativity, existence of an
Apparent Horizon implies the appearance of a future Event Horizon
outside of it. Therefore, one can excise the region inside the Apparent
Horizon (that also includes the singularity) for the stability of
the computation since nothing will come out of that region in classical
physics. For modified gravity theories, an Apparent Horizon need not
be inside the Event Horizon (See the discussion and references in
\cite{Baumgarte}).

Our task in this work is to give a detailed definition of the Apparent
Horizon and some related concepts and apply it to slowly rotating
and moving initial data  which was recently given in \cite{altas-tekin-apparenthorizon}. The layout of this work is follow: in section $\lyxmathsym{\mbox{II}}$ we introduce the necessary tools for the defining equation of an Apparent Horizon as a co-dimension two spatial hypersurface in $n$ dimensions and use the ADM decomposition of the metric to arrive at an equation in local coordinates,
in section $\text{\mbox{III}}$ we consider a conformally flat initial data for $n=1+3$ dimensions for which the momentum constraints can be solved exactly following the Bowen-York construction \cite{BY}; and we solve the Hamiltonian constraint for slowly moving and spinning initial data and compute the properties of the Apparent Horizon. In the Appendix we expound upon some technical points alluded to in the text.

\section{Derivation of the apparent horizon equation }

As stated above, the Event Horizon of a black hole, as a null hypersurface, cannot be determined
locally: one has to know the total spacetime to define it. On the
other hand, the Apparent Horizon can be determined locally in time. For this
purpose, we need to define a congruence of null geodesics and its
expansion. Our notations will be similar to those of the excellent lecture notes \cite{Eric0,Eric}.

As shown in Figure 1,  we have an $n$ dimensional spacetime manifold $\mathscr{M}$, with
a co-dimension one spacelike hypersurface $\Sigma$, that is $\dim\text{\ensuremath{\Sigma}}=n-1$;
and we introduce a co-dimension two subspace ${\mathcal{S}}$, $\dim{\mathcal{S}}=n-2$.
Let $n^{\mu}$ be a timelike unit vector orthogonal to $\Sigma$:
\begin{equation}
n^{\mu}n_{\mu}=-1,
\end{equation}
and $s^{\mu}$ be a spacelike unit vector orthogonal to ${\mathcal{S}}$
\begin{equation}
s^{\mu}s_{\mu}=1.
\end{equation}
We impose the condition that $n$ and $s$-vectors are perpendicular
to each other 
\begin{equation}
n^{\mu}s_{\mu}=0.
\end{equation}
Instead of these two vectors, one can also work with the ingoing null
vector $k^{\mu}$ and the outgoing null vector $\ell^{\mu}$, defined
respectively as follows (see Figure 2)
\begin{equation}
k^{\mu}:=\frac{1}{2}\left(n^{\mu}-s^{\mu}\right),\hskip1cm \ell^{\mu}:=n^{\mu}+s^{\mu}.\label{nullout}
\end{equation}
\begin{figure}
\centering \includegraphics[width=0.45\linewidth]{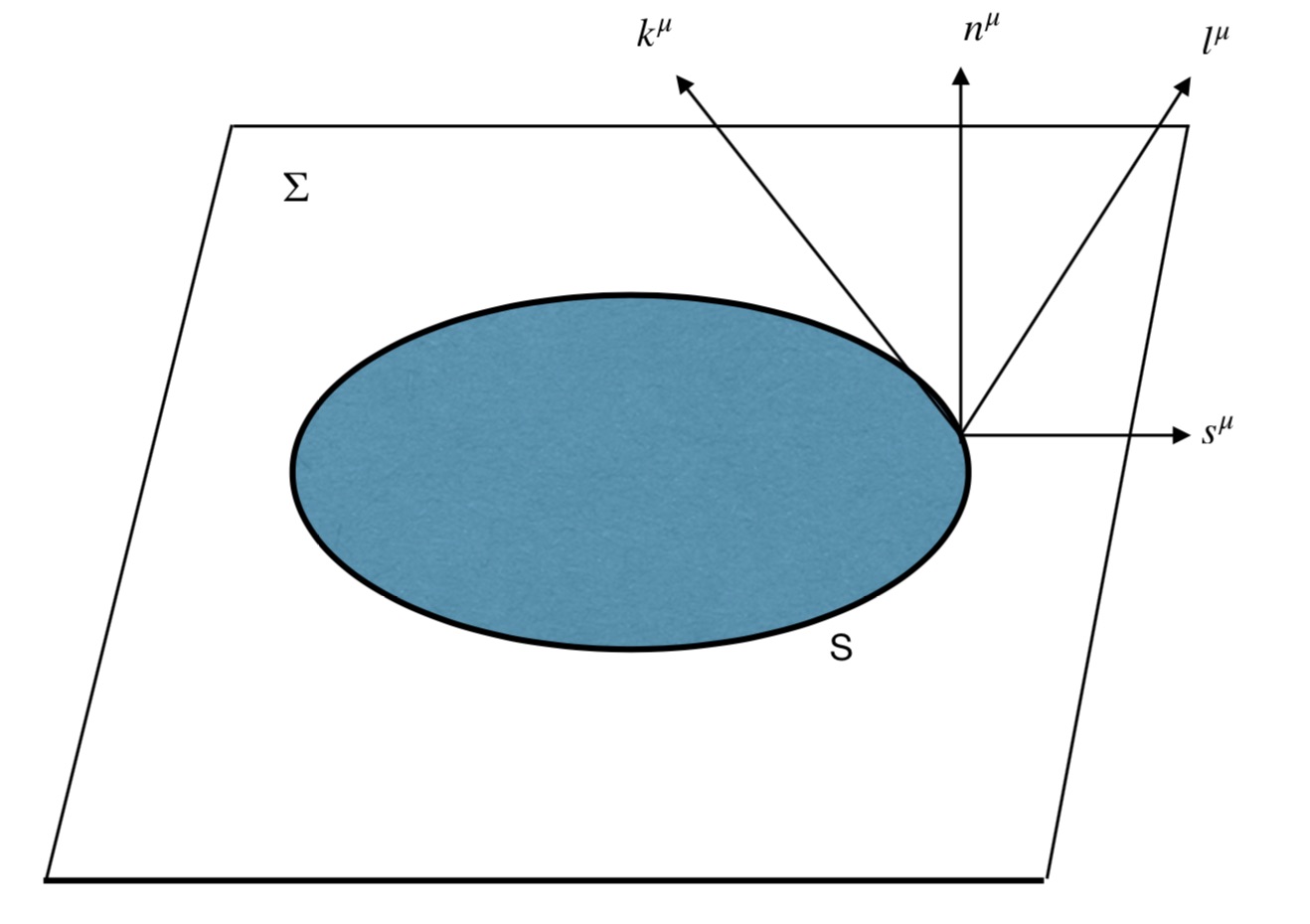}\caption{On an $n$ dimensional spacetime manifold $\mathscr{M}$, we introduce
a co-dimension one spacelike hypersurface $\Sigma$ and also a co-dimension
two subspace ${\mathcal{S}}$. Timelike unit vector $n^{\mu}$ is
orthogonal to $\Sigma$ and the spacelike unit vector $s^{\mu}$ is
orthogonal to ${\mathcal{S}}$. $k^{\mu}$ and $\ell^{\mu}$ denote the
ingoing and outgoing null vectors respectively.}
\label{subspaceblue} 
\end{figure}
The induced metric on the hypersurface $\Sigma$ is 
\begin{equation}
\gamma_{\mu\nu}=g_{\mu\nu}+n_{\mu}n_{\nu},\label{metriconsigma}
\end{equation}
while the induced metric on the subspace ${\mathcal{S}}$ reads 
\begin{equation}
q_{\mu\nu}=\gamma_{\mu\nu}-s_{\mu}s_{\nu}=g_{\mu\nu}+n_{\mu}n_{\nu}-s_{\mu}s_{\nu},\label{metriconS}
\end{equation}
where $\mu,\nu$ run over the spacetime directions.
The important concept here is the {\it extrinsic curvature } of both of these
surfaces. For the hypersurface $\Sigma$, we have 
\begin{equation}
K_{\mu\nu}:=-\gamma_{\mu\sigma}\gamma_{\nu\rho}\nabla^{\sigma}n^{\rho},\label{ex1}
\end{equation}
where $\nabla_{\mu}$ denotes the covariant derivative compatible
with the spacetime metric, $\nabla_{\mu}g_{\nu\rho}=0$. From a
more geometric vantage point, our definition is as follows: given
two vectors $(X,Y)$ on the tangent space at the point $p$, that is $T_{p}\Sigma$,
and $n$ being the unit normal to $\Sigma$, then the extrinsic curvature
of $\Sigma$ is  defined as $K(X,Y):=-\gamma(\nabla_{X}n,Y)$. So in local coordinates,
one can take $X=\partial_{\mu}$, $Y=\partial_{\nu}$ to get $K_{\mu\nu}:=K(\partial_{\mu},\partial_{\nu})=-\gamma(\nabla_{\partial_{\mu}}n,\partial_{\nu})$
which matches (\ref{ex1}). The minus sign is a convention. Equivalently, one has\footnote{One can also define the acceleration as $a :=\nabla_{n}n$.}
\begin{equation}
K_{\mu\nu}=-\nabla_{\mu}n_{\nu}-n_{\mu}n^{\sigma}\nabla_{\sigma}n_{\nu}.\label{extrinsic curvarure1}
\end{equation}
Similarly, we define the extrinsic curvature of the ($n-2$)-dimensional
space ${\mathcal{S}}$ as 
\begin{equation}
k_{\mu\nu}:=-q_{\mu\sigma}q_{\nu\rho}\nabla^{\sigma}s^{\rho},
\end{equation}
and using the definition of the induced metric (\ref{metriconS})
one obtains 
\begin{equation}
k_{\mu\nu}=-\nabla_{\mu}s_{\nu}-n_{\nu}n^{\sigma}\nabla_{\mu}s_{\sigma}-n_{\mu}n^{\sigma}\nabla_{\sigma}s_{\nu}-n_{\mu}n_{\nu}n_{\sigma}n_{\rho}\nabla^{\sigma}s^{\rho}+s_{\mu}s^{\sigma}\nabla_{\sigma}s_{\nu}+s_{\mu}n_{\nu}s_{\sigma}n_{\rho}\nabla^{\sigma}s^{\rho}.
\end{equation}
\begin{figure}
\centering \includegraphics[width=0.6\linewidth]{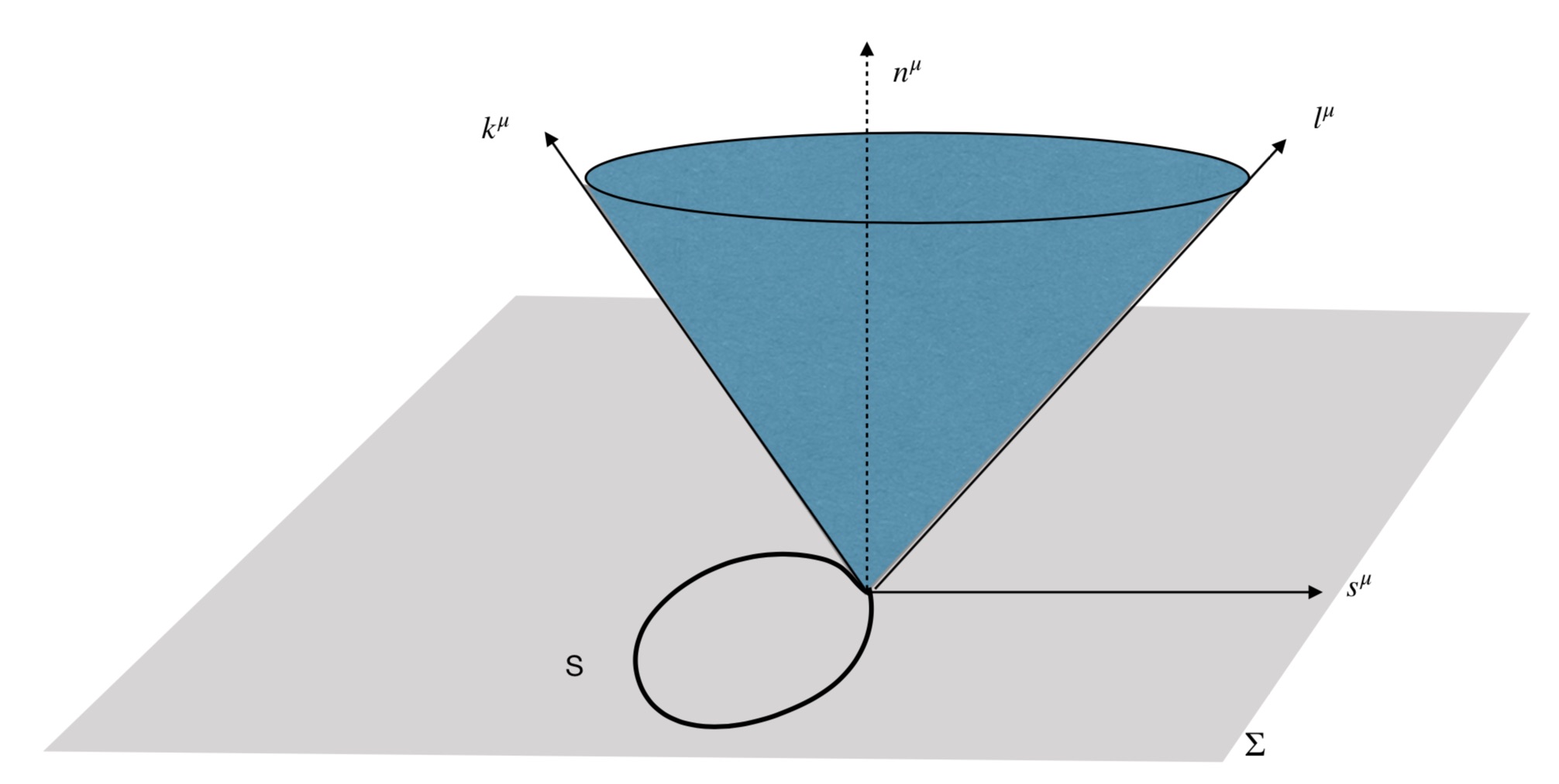}\caption{The unit vectors $n^{\mu}$, $s^{\mu}$ together with
the ingoing null vector $k^{\mu}$ and outgoing null vector $\ell^{\mu}$
are shown.}
\label{coneblue} 
\end{figure}
One defines the \textit{expansion } of the out-going null geodesic
congruence as 
\begin{equation}
\Theta_{\left(\ell\right)}:=q^{\mu\nu}\nabla_{\mu}\ell_{\nu},\label{nullgeodesic congruence equation}
\end{equation}
which is the divergence of the null geodesic congruence along its
propagation in the outgoing null direction. Using (\ref{nullout})
and the extrinsic curvatures of the hypersurface and the surface,
we can recast the expansion of the null geodesic congruence as 
\begin{equation}
\Theta_{\left(\ell\right)}=K+k+(n^{\mu}n^{\nu}-s^{\mu}s^{\nu})\left(K_{\mu\nu}+k_{\mu\nu}\right).
\end{equation}
Since $n^{\mu}n^{\nu}K_{\mu\nu}=0=n^{\mu}n^{\nu}k_{\mu\nu}=s^{\mu}s^{\nu}k_{\mu\nu}$,
$\Theta_{\left(\ell\right)}$ reduces to the following neat equation
as 
\begin{equation}
\Theta_{\left(\ell\right)}=-K-k+s^{\mu}s^{\nu}K_{\mu\nu}.
\end{equation}
Equivalently one has 
\begin{equation}
\Theta_{\left(\ell\right)}=-q^{\mu\nu}\Bigl(K_{\mu\nu}+k_{\mu\nu}\Bigr),
\end{equation}
or 
\begin{equation}
\Theta_{\left(\ell\right)}=-q^{ij}\Bigl(K_{ij}+k_{ij}\Bigr),
\end{equation}
where the $i,j$ indices run over coordinates on the hypersurface
$\Sigma$.

The expansion $\Theta_{\left(\ell\right)}$ is employed to define the very important
concept of a \textit{trapped surface}. An outer trapped surface on
$\Sigma$ is a \textit{closed } (that is compact without a boundary)
co-dimension two  surface such that for outgoing null geodesics orthogonal
to the surface, one has $\Theta_{\left(\ell\right)}<0$ {\it everywhere} on
the surface. The subset of $\Sigma$ that contains the trapped surfaces
is called the \textit{trapped region} ${\mathcal{T}}$, a co-dimension one surface. Finally, Apparent
Horizon is the boundary of the trapped region (an obviously spatial
surface) which we shall denote by ${\mathcal{S}}:=\partial{\mathcal{T}}$.
By definition Apparent Horizon is a marginally outer trapped surface
(MOTS) and satisfies the  Apparent Horizon equation: 
\begin{equation}
\Theta_{\left(\ell\right)}=-K-k+s^{\mu}s^{\nu}K_{\mu\nu}=0.\label{apparenthorizonequation-version2}
\end{equation}
It is clear that for the case of time-symmetric initial data ($K_{\mu\nu}=0$),
the Apparent Horizon becomes a minimal surface since $k=0$.

Now that we have defined the Apparent Horizon, given a metric in some
coordinates, to proceed we need to lay out in detail how (\ref{apparenthorizonequation-version2})
is expressed in terms of the metric functions. For this purpose we
choose the ADM decomposition of the metric \cite{ADM}.

Let $N=N(t,x^{i})$ be the lapse function and $N^{i}=N^{i}(t,x^{j})$
be the shift vector, then the line-element reads 
\begin{equation}
ds^{2}=(N_{i}N^{i}-N^{2})dt^{2}+2N_{i}dt\thinspace dx^{i}+\gamma_{ij}dx^{i}\thinspace dx^{j},
\end{equation}
or in components one has 
\begin{equation}
g_{00}=N_{i}N^{i}-N^{2},~~\ ~~\ ~~\ ~~\ g_{0i}=N_{i},~~\ ~~\ ~~\ ~~\ g_{ij}=\gamma_{ij},
\end{equation}
with the inverses given as 
\begin{equation}
g^{00}=-N^{-2},~~\ ~~\ ~~\ ~~\ g^{0i}=N^{i}N^{-2},~~\ ~~\ ~~\ ~~\ g^{ij}=\gamma^{ij}-N^{i}N^{j}N^{-2}.
\end{equation}
Using the definition (\ref{metriconsigma}), one has 
\begin{equation}
g_{ij}=\gamma_{ij}-n_{i}n_{j}=\gamma_{ij},
\end{equation}
hence $n_{i}=0$. Similarly the relation $g^{ij}=\gamma^{ij}-n^{i}n^{j}=\gamma^{ij}-N^{i}N^{j}N^{-2}$
yields $n^{i}=\pm N^{i}/N$. Since $n^{\mu}$ is a timelike vector,
using $n_{i}=0$, one has $n_{0}=\pm N$ and choosing for $N>0$,
we choose the plus sign for the future-directed time-like vector to
arrive at 
\begin{equation}
n^{\mu}=\left(\frac{1}{N},-\frac{N^{i}}{N}\right),\hskip1cmn_{\mu}=(-N,\vec{0}).
\end{equation}
We can work out the additional relations between the spacetime metric
$g$ and the metric of the hypersurface $\gamma$ as follows 
\begin{equation}
g_{00}=\gamma_{00}-n_{0}n_{0}=N_{i}N^{i}-N^{2},
\end{equation}
which yields $\gamma_{00}=N_{i}N^{i}$. And similarly 
\begin{equation}
g_{0i}=\gamma_{0i}-n_{0}n_{i}=N_{i}
\end{equation}
yields $\gamma_{0i}=N_{i}$; and from the inverse metric relations,
one obtains $\gamma^{0\mu}=0$.

Similar computations for the co-dimension 2 spatial subspace ${\mathcal{S}}$,
after using the condition $n_{\mu}s^{\mu}=0$, yield $s^{0}=0$ and
$q^{0\mu}=0$.

Now we go back to (\ref{nullgeodesic congruence equation}) and express
it for the apparent horizon as 
\begin{equation}
q^{ij}\left(\nabla_{i}n_{j}+\nabla_{i}s_{j}\right)=0.
\end{equation}
From (\ref{extrinsic curvarure1}), one has $K_{ij}=-\nabla_{i}n_{j}=-N\Gamma_{ij}^{0}$
and we obtain\footnote{ Note that, explicitly, we have $K_{ij}=\frac{1}{2N}\left(D_{i}N_{j}+D_{j}N_{i}-\partial_{t}\gamma_{ij}\right)$.}
\begin{equation}
q^{ij}\left(K_{ij}-\partial_{i}s_{j}+\Gamma_{ij}^{0}s_{0}+\Gamma_{ij}^{k}s_{k}\right)=0.
\end{equation}
We denote the Christoffel connection of the induced metric $\gamma$
as $^{\Sigma}\Gamma_{ij}^{k}$. Then substituting the corresponding
components of the Christoffel connection one has 
\begin{equation}
\Gamma_{ij}^{k}={}^{\Sigma}\Gamma_{ij}^{k}+\frac{N^{k}}{N}K_{ij},
\end{equation}
and we arrive at 
\begin{equation}
q^{ij}\left(K_{ij}-D_{i}s_{j}+\frac{1}{N}K_{ij}\left(N^{k}s_{k}-s_{0}\right)\right)=0,
\end{equation}
where $D_{i}$ denotes the covariant derivative compatible with the
spatial metric $\gamma$, $D_{i}\gamma_{jk}=0$. One has $N^{k}s_{k}-s_{0}=0$,
so then the equation defining the apparent horizon becomes 
\begin{equation}
q^{ij}\left(K_{ij}-D_{i}s_{j}\right)=0.\label{final_AH}
\end{equation}

Before we start working out an example, let us note that there
is another simple expression of the $\theta_{(\ell)}$ in (\ref{nullgeodesic congruence equation})
and hence equation (\ref{final_AH}). One can show that (see section A of the Appendix
for the proof) Lie-dragging the metric on ${\mathcal{S}}$ along $\ell$
yields exactly the expansion: namely, one has 
\begin{equation}
\theta_{(\ell)}=q^{\mu\nu}\nabla_{\mu}l_{\nu}=\frac{1}{2}q^{\mu\nu}{\mathcal{L}}_{\ell}q_{\mu\nu},\label{proof1}
\end{equation}
where ${\mathcal{L}}_{\ell}$ denotes the Lie-derivative along the
vector $\ell$.  In section B of the Appendix, $\theta_{(\ell)}$ is derived from the minimization of the area along the outing null direction which also yields a complementary physical picture. 

\section{Apparent horizon detection}

\subsection{ The equation in explicit form}

From now on we shall work in $n=1+3$ dimensions. Assume now that
the local coordinates on $\Sigma$ are denoted as $(r,\theta,\phi)$
and that the location of the Apparent Horizon depends both on $\theta$
and $\phi$. The equation to be solved is 
\begin{equation}
q^{ij}\left(\partial_{i}s_{j}-^{\Sigma}\Gamma_{ij}^{k}s_{k}-K_{ij}\right)=0.\label{equationinaxisymmetry0}
\end{equation}
Assume that the surface ${\mathcal{S}}$ can be parameterized as a
level set such that 
\begin{equation}
\Phi(r,\theta,\phi):=r-h(\theta,\phi)=0,
\end{equation}
with $h$ being a sufficiently differentiable function of its arguments.
Since $s^{i}$ is normal to the surface, one has $s_{i}\sim\partial_{i}\varPhi$;
and because it is a normal vector, let $s_{i}:=\lambda\partial_{i}\varPhi$
and define $m_{i}:=\partial_{i}\varPhi$, which yields 
\begin{equation}
s_{i}=\lambda\Big(1,-\partial_{\theta}h,-\partial_{\phi}h\Big).
\end{equation}
To proceed further, let us take the metric on $\Sigma$ to be conformally
flat as in \cite{BY} 
\begin{equation}
\gamma_{ij}=\psi^{4}\begin{pmatrix}1 & 0 & 0\\
0 & r^{2} & 0\\
0 & 0 & r^{2}\sin^{2}\theta
\end{pmatrix},
\end{equation}
then one has 
\begin{equation}
s^{i}=\lambda\left(\gamma^{rr},-\gamma^{\theta\theta}\partial_{\theta}h,-\gamma^{\phi\phi}\partial_{\phi}h\right),
\end{equation}
with the normalization factor given as 
\begin{equation}
\lambda=\Big(\gamma^{rr}+\gamma^{\theta\theta}(\partial_{\theta}h)^{2}+\gamma^{\phi\phi}(\partial_{\phi}h)^{2}\Big)^{-1/2}.
\end{equation}
As should be clear at this stage, the Apparent Horizon equation will
be a rather complicated non-linear partial differential equation with little hope to yield an exact analytical solution.
Let us further assume (following \cite{BY}) $\gamma^{ij}K_{ij}=K=0$, which is called the maximal slicing gauge. 
Then (\ref{equationinaxisymmetry0}) reads more explicitly as 
\begin{equation}
\gamma^{ij}\partial_{i}m_{j}-\gamma^{ij}\Gamma_{ij}^{k}m_{k}-\lambda^{2}m^{i}m^{j}\partial_{i}m_{j}+\lambda^{2}m^{i}m^{j}m_{k}\Gamma_{ij}^{k}+\lambda m^{i}m^{j}K_{ij}=0.\label{apparenthorizonequation}
\end{equation}
After working out each piece in a somewhat tedious computation, one
arrives at 
\begin{eqnarray}
 &  & -\gamma^{\theta\theta}\partial_{\theta}^{2}h-\gamma^{\phi\phi}\partial_{\phi}^{2}h-\frac{1}{2}\Bigl((\gamma^{rr})^{2}\partial_{r}\gamma_{rr}-\gamma^{\theta\theta}\gamma^{rr}\partial_{r}\gamma_{\theta\theta}-\gamma^{\phi\phi}\gamma^{rr}\partial_{r}\gamma_{\phi\phi}+\partial_{\theta}h\gamma^{\phi\phi}\gamma^{\theta\theta}\partial_{\theta}\gamma_{\phi\phi}\Bigr)\nonumber \\
 &  & +\lambda^{2}\Bigl((\gamma^{\theta\theta})^{2}(\partial_{\theta}h)^{2}\partial_{\theta}^{2}h+(\gamma^{\phi\phi})^{2}(\partial_{\phi}h)^{2}\partial_{\phi}^{2}h+2\gamma^{\phi\phi}\gamma^{\theta\theta}\partial_{\phi}h\partial_{\theta}h\partial_{\theta}\partial_{\phi}h\Bigr)\nonumber \\
 &  & +\frac{\lambda^{2}}{2}\Bigl((\gamma^{rr})^{3}\partial_{r}\gamma_{rr}+(\gamma^{\theta\theta})^{2}\gamma^{rr}(\partial_{\theta}h)^{2}\partial_{r}\gamma_{\theta\theta}+(\gamma^{\phi\phi})^{2}\gamma^{rr}(\partial_{\phi}h)^{2}\partial_{r}\gamma_{\phi\phi}\nonumber \\
 &  & ~~~~~~~~~~~~~-(\partial_{\phi}h)^{2}\partial_{\theta}h(\gamma^{\phi\phi})^{2}\gamma^{\theta\theta}\partial_{\theta}\gamma_{\phi\phi}\Bigr)\nonumber \\
 &  & +\lambda\Bigl((\gamma^{rr})^{2}K_{rr}+(\gamma^{\theta\theta})^{2}(\partial_{\theta}h)^{2}K_{\theta\theta}+(\gamma^{\phi\phi})^{2}(\partial_{\phi}h)^{2}K_{\phi\phi}-2\gamma^{rr}\gamma^{\theta\theta}\partial_{\theta}hK_{r\theta}\nonumber \\
 &  & ~~~~~~~~~~~~~~-2\gamma^{rr}\gamma^{\phi\phi}\partial_{\phi}hK_{r\phi}+2\gamma^{\theta\theta}\gamma^{\phi\phi}\partial_{\theta}h\partial_{\phi}hK_{\theta\phi}\Bigr)=0.\label{yarab}
\end{eqnarray}
Given the metric $\gamma_{ij}$ and the extrinsic curvature $K_{ij}$,
one can find numerical solutions of this equation up to the desired
accuracy. Our goal here is to find approximate analytical solutions
to some physically reasonable initial data which must satisfy the
Hamiltonian and the momentum constraints on the hypersurface $\Sigma$
which we discuss next.

\subsection{1+3 form of Einstein equations}

The Hamiltonian and the momentum constraints on the hypersurface $\Sigma$
follow from Einstein's equations as 
\begin{eqnarray}
 &  & -^{\Sigma}R-K^{2}+K_{ij}K^{ij}-2\kappa T_{nn}=0,\nonumber \\
 &  & 2D_{k}K_{i}^{k}-2D_{i}K-2\kappa T_{ni}=0.~~~~~~~~~~~~~~\label{Einstein_c}
\end{eqnarray}
We chosen $K=0$ and consider the vacuum case with $T_{\mu\nu}=0$.
Of course this initial data evolves in time and the remaining parts
of the Einstein equations written as a dynamical system are given
as 
\begin{equation}
\frac{\partial}{\partial t}\gamma_{ij}=-2NK_{ij}+D_{i}N_{j}+D_{j}N_{i},
\end{equation}
\begin{equation}
\frac{\partial}{\partial t}K_{ij}=-N\left(R_{ij}-{}^{\Sigma}R_{ij}-KK_{ij}+2K_{ik}K_{j}^{k}\right)+\mathscr{L}_{\vec{N}}K_{ij}-D_{i}D_{j}N,
\end{equation}
where $\text{\ensuremath{\mathscr{L}}}_{\vec{N}}$ is the Lie derivative
along the shift vector $N^{i}$.  Derivation of these well-known equations can be found in many textbooks, see our derivation in \cite{our_dain_paper}.

\subsection{Conformally flat Bowen-York type data}

For a conformally flat hypersurface $\Sigma$ ($\gamma_{\ij}=\psi^{4}f_{ij}$
with $f$ being the flat metric in some coordinates), the constraint
equations (\ref{Einstein_c}) (together with the "maximal slicing"
condition $K=0$) reduce to a non-linear elliptic equation and an
easily solvable linear equation, respectively given as 
\begin{eqnarray}
 &  & \hat{D}_{i}\hat{D}^{i}\psi=-\frac{1}{8}\psi^{-7}\hat{K}_{ij}^{2},\label{elliptic0}\\
 &  & \hat{D}^{i}\hat{K}_{ij}=0,\label{trans}
\end{eqnarray}
with $\hat{D}_{i}f_{jk}=0$ and $K_{ij}=\psi^{-2}\hat{K}_{ij}$.

Bowen and York \cite{BY} gave the following 7-parameter ($p_{i},a,{\cal {J}}_{i}$)
solution to (\ref{trans}) on $\mathbb{R}^{3}$ whose origin is removed:
\begin{eqnarray}
 &  & \hat{K}_{ij}=\frac{3}{2r^{2}}\Big(p_{i}n_{j}+p_{j}n_{i}+(n_{i}n_{j}-f_{ij})p\cdot n\Big)+\epsilon\frac{3a^{2}}{2r^{4}}\Big(p_{i}n_{j}+p_{j}n_{i}+(f_{ij}-5n_{i}n_{j})p\cdot n\Big)\nonumber \\
 &  & ~~~~~~~~+\frac{3}{r^{3}}{\cal {J}}^{l}n^{k}\Big(\varepsilon_{kil}n_{j}+\varepsilon_{kjl}n_{i}\Big),\label{Bowen-York_extrinsiccurvature}
\end{eqnarray}
where $r>0$ is the radial coordinate, $n^{i}$ is the unit normal
on a sphere of radius $r$ (not related to the unit normal to $\Sigma$);
$\epsilon=\pm1$ and $p\cdot n=p^{k}n_{k}$. At this stage, one should
note that the physical meaning of the parameters ($p_{i},a,{\cal {J}}_{i}$)
is not clear; secondly, linearity of (\ref{trans}) means that each
bracketed term solves the equation separately. For the sake of simplicity,
we shall choose $a=0$ in what follows.

Here we follow \cite{altas-tekin-apparenthorizon}. We shall need the following expression for the right-hand side of
(\ref{elliptic0}) 
\begin{equation}
\hat{K}_{ij}\hat{K}^{ij}=\frac{9}{2r^{4}}\left(p^{2}+2(\vec{p}\cdot\vec{n})^{2}\right)+\frac{18}{r^{5}}\left(\vec{J}\times\vec{n}\right)\cdotp\vec{p}+\frac{18}{r^{6}}\left(\vec{J}\times\vec{n}\right)\cdotp\left(\vec{J}\times\vec{n}\right).\label{kare}
\end{equation}
Inserting this expression to (\ref{elliptic0}), one arrives at the complicated Hamiltonian constraint which can only be solved exactly after making several assumptions. We shall not go into that discussion which was given in \cite{Altas} in some detail. 

\subsection{Conserved quantities }

To understand the physical meaning of the parameters in the solution, 
we shall assume that the spacetime is asymptotically flat, hence the
conformal factor behaves as 
\begin{equation}
\psi(r)=1+{\mathcal{O}}(1/r),\hskip1cm{\text{as}}\,\,\,r\to\infty.
\end{equation}
Then one has the conserved {\it total momentum} associated to $\Sigma$
easily written as a boundary integral on a sphere at spatial infinity:
\begin{equation}
P_{i}=\frac{1}{8\pi}\int_{S_{\infty}^{2}}dS\,n^{j}\,K_{ij}=\frac{1}{8\pi}\int_{S_{\infty}^{2}}dS\,n^{j}\,\hat{K}_{ij}.\label{mom}
\end{equation}
Observe, from the second equality, that only the leading term  in the conformal factor is relevant for this and the following computation.
The total conserved {\it total angular momentum} can also be found easily as 
\begin{equation}
J_{i}=\frac{1}{8\pi}\varepsilon_{ijk}\int_{S_{\infty}^{2}}dS\,n_{l}\,x^{j}K^{kl}=\frac{1}{8\pi}\varepsilon_{ijk}\int_{S_{\infty}^{2}}dS\,n_{l}\,x^{j}\hat{K}^{kl}.\label{dad}
\end{equation}
Given (\ref{mom}) and (\ref{dad}), it is straightforward to compute
the integrals for the extrinsic curvature (\ref{Bowen-York_extrinsiccurvature})
which at the end yield $P_{i}=p_{i}$ and $J_{i}={\cal {J}}_{i}$.
So for the computation of these two quantities, let us note once again that, the full form of the conformal
factor is not needed; one only needs to know its behavior at infinity,
that is the ${\mathcal{O}}(1)$ term.

From these two conserved quantities, one can see that physically the
assumed extrinsic curvature (\ref{Bowen-York_extrinsiccurvature})
belongs to a a self-gravitating system (a curved vacuum) with non-zero
momentum and angular momentum. To compute the total mass-energy, the
ADM energy, of the system, the ${\mathcal{O}}(1)$ term of the conformal
factor is not sufficient. For that computation we keep the next
order term and assume 
\begin{equation}
\psi(r)=1+\frac{E}{2r}+{\mathcal{O}}(1/r^{2})\hskip1cm{\text{as}}\,\,\,r\to\infty.\label{asymp1}
\end{equation}
Then defining the deviation from the background as $h_{ij} :=(\psi^{4}-1)\delta_{ij}$, the ADM energy simplifies
as 
\begin{equation}
E_{ADM}=\frac{1}{16\pi}\int_{S_{\infty}^{2}}dS\,n_{i}\,\Big(\partial_{j}h^{ij}-\partial_{i}h_{j}^{j}\Big)=-\frac{1}{2\pi}\int_{S_{\infty}^{2}}dS\,n^{i}\,\partial_{i}\psi,
\end{equation}
whose explicit evaluation for (\ref{asymp1}) yields $E_{ADM}=E$, which of course at this stage is almost a tautology: we  have to find the constant $E$ by solving the Hamiltonian
constraint.

\subsection{Approximate solution of the Hamiltonian constraint for a boosted
slowly rotating gravitating system}

To solve the elliptic equation (\ref{elliptic0}) using (\ref{kare}),
let us take $\hat{k}$ to be the direction of the conserved angular
momentum and choose $\vec{p}$ to be in the $xz$ plane (this is just
a choice of the orientation of the coordinates and no generality is
lost) 
\begin{equation}
\vec{J}=J\hat{k},\hskip1cm\vec{p}=p\sin\theta_{0}\hat{i}+p\cos\theta_{0}\hat{k},
\end{equation}
with $\theta_{0}$ a fixed, conserved angle. Then the Hamiltonian constraint
(\ref{elliptic0}) becomes 
\begin{equation}
\hat{D}_{i}\hat{D}^{i}\psi=\psi^{-7}\left(\frac{9Jp}{4r^{5}}c_{1}\sin\theta\sin\phi-\frac{9J^{2}}{4r^{6}}\sin^{2}\theta-\frac{9p^{2}}{16r^{4}}(1+2(c_{1}\sin\theta\cos\phi+c_{2}\cos\theta)^{2})\right),\label{hamiltoian constraint}
\end{equation}
where $c_{1}:=\sin\theta_{0},c_{2}:=\cos\theta_{0}.$

Needless to say, an exact solution of this equation is hopeless, therefore
we shall search for the lowest order perturbative solution assuming
an expansion in terms of the momentum and spin which corresponds to
a curved 3-surface with a small linear and small angular momentum. In \cite{Gleiser} the slowly spinning case with no linear momentum was solved in the leading order; and in \cite{Dennison-Baumgarte} slowly moving without spin was solved and in \cite{altas-tekin-apparenthorizon}, both motions were considered at the leading order. We now present this solution.

A cursory inspection on the right-hand side suggests that one should
have a double series of the form 
\begin{equation}
\psi(r,\theta,\phi):=\psi^{(0)}+J^{2}\psi^{(J)}+p^{2}\psi^{(p)}+Jp\psi^{(Jp)}+\mathcal{O}(p^{4},J^{4},p^{2}J^{2}),\label{expansionof psi}
\end{equation}
where the functions on the right-hand side depend on $(r,\theta,\phi)$.
At the zeroth order, one has the usual Laplace equation 
\begin{equation}
\hat{D}_{i}\hat{D}^{i}\psi^{(0)}=0,
\end{equation}
which needs boundary conditions to be uniquely solved. The following
boundary conditions as employed by \cite{Dennison-Baumgarte} are
apt for the problem at hand: at spatial infinity one demands 
\begin{equation}
\lim_{r\rightarrow\infty}\psi(r)=1,~~~~~~~~~\psi(r)>0\label{boundarycondition1}
\end{equation}
and near the origin one has 
\begin{equation}
\lim_{r\rightarrow\ 0}\psi(r)=\psi^{(0)},\label{boundarycondition2}
\end{equation}
where $\psi^{(0)}$ might have a singularity at the origin.  In fact the 
zeroth order solution satisfying these boundary conditions reads
\begin{equation}
\psi^{(0)}=1+\frac{a}{r},
\end{equation}
where $a$ is a constant of integration at this stage not to be confused
with the one in (\ref{Bowen-York_extrinsiccurvature}). The equations
at the next order are 
\begin{eqnarray}
 &  & \hat{D}_{i}\hat{D}^{i}\psi^{(J)}=-\frac{9}{4}\sin^{2}\theta\frac{r}{(r+a)^{7}},\label{Jequation}\\
 &  & \hat{D}_{i}\hat{D}^{i}\psi^{(Jp)}=\frac{9}{4}c_{1}\sin\theta\sin\phi\frac{r^{2}}{(r+a)^{7}},\label{Jpequation}\\
 &  & \hat{D}_{i}\hat{D}^{i}\psi^{(p)}=-\frac{9}{16}\left(1+2(c_{1}\sin\theta\cos\phi+c_{2}\cos\theta)^{2}\right)\frac{r^{3}}{(r+a)^{7}}.\label{pequation}
\end{eqnarray}
These are linear equations whose solutions can be found with the help
of the following spherical harmonics : 
\begin{eqnarray*}
 &  & Y_{0}^{0}(\theta,\phi)=\frac{1}{\sqrt{4\pi}},~~~~~~~~~~~~~~~Y_{1}^{0}(\theta,\phi)=\sqrt{\frac{3}{4\pi}}\cos\theta,~~~~~~Y_{2}^{0}(\theta,\phi)=\sqrt{\frac{5}{16\pi}}(3\cos^{2}\theta-1),\\
 &  & Y_{1}^{-1}(\theta,\phi)=\sqrt{\frac{3}{4\pi}}\sin\theta\sin\phi,~~Y_{2}^{1}(\theta,\phi)=\sqrt{\frac{15}{4\pi}}\sin\theta\cos\theta\cos\phi,~~Y_{1}^{1}(\theta,\phi)=\sqrt{\frac{3}{4\pi}}\sin\theta\cos\phi.
\end{eqnarray*}
Then a close inspection of (\ref{Jequation}) suggests that the proper
ansatz for $\psi^{(J)}$ should be of the form 
\[
\psi^{(J)}(r,\theta,\phi)=\psi_{0}^{(J)}(r)Y_{0}^{0}(\theta,\phi)+\psi_{1}^{(J)}(r)Y_{2}^{0}(\theta,\phi),
\]
from which the solution obeying the boundary conditions (\ref{boundarycondition1},
\ref{boundarycondition2}) can be found to be 
\begin{equation}
\psi^{(J)}(r,\theta,\phi)=\frac{\left(a^{4}+5a^{3}r+10a^{2}r^{2}+5ar^{3}+r^{4}\right)}{40a^{3}(a+r)^{5}}-\frac{r^{2}}{40a(a+r)^{5}}(3\cos^{2}\theta-1).
\end{equation}
To solve (\ref{Jpequation}) one should take 
\[
\psi^{(Jp)}(r,\theta,\phi)=\psi_{0}^{(Jp)}(r)Y_{0}^{0}(\theta,\phi)+\psi_{1}^{(Jp)}(r)Y_{1}^{-1}(\theta,\phi),
\]
for which the solution obeying the boundary conditions is 
\begin{equation}
\psi^{(Jp)}(r,\theta,\phi)=-\frac{c_{1}r\left(a^{2}+5ar+10r^{2}\right)}{80a(a+r)^{5}}\sin\theta\sin\phi.
\end{equation}
The $\psi^{(p)}$ equation (\ref{pequation}) is similar albeit slightly
more complicated: the proper ansatz reads 
\[
\psi^{(p)}=\psi_{0}^{(p)}(r)Y_{0}^{0}(\theta,\phi)+\psi_{1}^{(p)}(r)Y_{1}^{1}(\theta,\phi)^{2}+\psi_{2}^{(p)}(r)Y_{2}^{1}(\theta,\phi)+\psi_{3}^{(p)}(r)Y_{1}^{0}(\theta,\phi)^{2},
\]
from which four equations follow whose solutions are as follows: 
\begin{eqnarray}
\psi_{0}^{(p)}(r) & = & -\frac{\sqrt{\pi}\left(84a^{6}+378a^{5}r+653a^{4}r^{2}+514a^{3}r^{3}+142a^{2}r^{4}-35ar^{5}-25r^{6}\right)}{80ar^{2}(a+r)^{5}}\nonumber \\
 &  & -\frac{21\sqrt{\pi}a}{20r^{3}}\log\frac{a}{a+r},
\end{eqnarray}
and

\begin{eqnarray}
\psi_{1}^{(p)}(r) & = & \frac{\pi c_{1}^{2}\left(84a^{5}+378a^{4}r+658a^{3}r^{2}+539a^{2}r^{3}+192ar^{4}+15r^{5}\right)}{40r^{2}(a+r)^{5}}\nonumber \\
 &  & +\frac{21\pi ac_{1}^{2}}{10r^{3}}\log\frac{a}{r+a}.\label{psi_1^p(r)}
\end{eqnarray}
$\psi_{2}^{(p)}(r)$ can be obtained from (\ref{psi_1^p(r)}) with
the replacement $c_{1}^{2}\rightarrow\sqrt{\frac{3}{5\pi}}c_{1}c_{2}$
and $\psi_{3}^{(p)}(r)$ can be obtained from (\ref{psi_1^p(r)})
with the replacement $c_{1}^{2}\rightarrow c_{2}^{2}$ . All these pieces can be combined
to get $\psi^{(p)}$ at this stage, but a depiction of the final result
is redundant since all the parts are given above and the final expression
is cumbersome. We have now all the information at our disposal to
compute the relevant quantities defined on $\Sigma$ including the
location of the Apparent Horizon.

First let us revisit the ADM energy computation which we started
above: We need the dominant terms up to and including $\mathcal{O}(\frac{1}{r})$
in $\psi(r,\theta,\phi)$. A quick power series expansion yields 
\begin{equation}
\psi(r)=1+\frac{a}{r}+\frac{J^{2}}{40a^{3}r}+\frac{5p^{2}}{32ar}+\mathcal{O}(\frac{1}{r^{2}}),\label{solution}
\end{equation}
in which the $Jp$ term appears at $\mathcal{O}(\frac{1}{r^{2}})$
and therefore makes no contribution to the energy. Then from (\ref{asymp1}),
the ADM energy of the solution follows as 
\begin{equation}
E_{\text{ADM}}=2a+\frac{J^{2}}{20a^{3}}+\frac{5p^{2}}{16a}.
\end{equation}
So one can immediately see that for vanishing spin and vanishing linear
momentum (that is the case of the Schwarschild black hole written in the isotropic coordinates) the constant $a$ is related to the mass of the Schwarschild
black hole mass as $a=M/2$.

\subsection{Apparent Horizion area and the irreducible mass}

While studying the efficient processes of extracting energy from rotating
black holes, Christodoulou \cite{Chris} realized\footnote{Note that after Hawking's area theorem \cite{hawk2} which came later
than Christodoulou's observation, it became clear that there must
be an irreducible mass at the classical level.} that there is an irreducible mass $M_{\text{irr}}$ which is related
to the area $A_{\text{EH}}$ of a \textit{section} of the event horizon
via 
\begin{equation}
M_{\text{irr}}:=\sqrt{\frac{A_{\text{EH}}}{16\pi}}.\label{irreduciblemassformula}
\end{equation}
For a moving, rotating black hole, the total energy was obtained in
\cite{Chris} as 
\begin{equation}
E^{2}=M_{\text{irr}}^{2}+p^{2}+\frac{J^{2}}{4M_{\text{irr}}^{2}},\label{chris2}
\end{equation}
in which the physical meaning of each part is clear.

Since we have a dynamical, evolving system, we have at our disposal
the area of the Apparent Horizon, not a section of the Event Horizon.
But, following \cite{Dennison-Baumgarte}, a good approximation to
$M_{\text{irr}}$ can be given with the help of the area of the Apparent
Horizon via 
\begin{equation}
M_{\text{irr}}:=\sqrt{\frac{A_{\text{AH}}}{16\pi}}.\label{irreduciblemassformula}
\end{equation}
As we shall see, this definition yields the correct expression for
the energy of our system obtained from an expansion of (\ref{chris2}).
But first we need to find the location of the Apparent Horizon, namely
solve (\ref{yarab}) up to the accuracy we have been working with.
That area is given simply as 
\begin{equation}
A=\intop_{0}^{2\pi}d\phi\intop_{0}^{\pi}d\theta\sqrt{\det q},
\end{equation}
which yields the following  {\it exact} form:
\begin{equation}
A_{\text{AH}}=\intop_{0}^{2\pi}d\phi\intop_{0}^{\pi}d\theta\thinspace\sin\theta\thinspace\psi^{4}\thinspace h^{2}\left(1+\frac{1}{h^{2}}\left(\partial_{\theta}h\right)^{2}+\frac{1}{h^{2}\sin^{2}\theta}\left(\partial_{\phi}h\right)^{2}\right)^{1/2}.\label{alan3}
\end{equation}
Hence to get the area, all we need is to find the location of the
Apparent Horizon up to first order in the spin and momentum. This
suggests the following ansatz: 
\begin{equation}
h(\theta,\phi)=h^{0}+ph^{p}+Jh^{J}+\mathcal{O}(p^{2},J^{2},Jp),
\end{equation}
where 
\begin{equation}
\partial_{r}h=0,~~~~~~\partial_{r}h^{0}=0=\partial_{\theta}h^{0}=\partial_{\phi}h^{0}.
\end{equation}
Ignoring the terms such as $(\partial_{\theta}h)^{2}$, $(\partial_{\phi}h)^{2}$
and $\partial_{\theta}h\partial_{\phi}h$, (\ref{yarab}) reduces
to 
\begin{equation}
\partial_{\theta}^{2}h+\frac{1}{\sin^{2}\theta}\partial_{\phi}^{2}h+\cot\theta\partial_{\theta}h-2r-4r^{2}\frac{\partial_{r}\psi}{\psi}+\frac{6J}{\psi^{4}r^{2}}\partial_{\phi}h-\frac{3p}{\psi^{4}}\Big(c_{1}\sin\theta\cos\phi+c_{2}\cos\theta\Big)=0.\label{result1}
\end{equation}
At the zeroth order, $\mathcal{O}(p^{0},J^{0})$, it yields 
\begin{equation}
1+2r\frac{\partial_{r}\psi}{\psi}=0,
\end{equation}
with $\psi=1+\frac{a}{r}$ ; setting $r=h$, one finds 
\begin{equation}
h^{0}=a.
\end{equation}
This solution identifies the parameter $a$ as the location of
the apparent horizon at the lowest, dominant, order. For example,
for the Schwarzschild black hole $h=2M$ (as noted above) would be
the exact solution for which the Event Horizon and the Apparent horizon
coincide in these conformally flat, isotropic coordinates.

At $\mathcal{O}(p)$ we have an inhomogeneous, linear Helmholtz equation
on a sphere ($S^{2}$), 
\begin{equation}
\Bigg(\partial_{\theta}^{2}+\frac{1}{\sin^{2}\theta}\partial_{\phi}^{2}+\cot\theta\partial_{\theta}-1\Bigg)h^{p}=\frac{3}{16}\Big(c_{1}\sin\theta\cos\phi+c_{2}\cos\theta\Big),
\end{equation}
while at $\mathcal{O}(J)$, we have a homogeneous one: 
\begin{equation}
\left(\partial_{\theta}^{2}+\frac{1}{\sin^{2}\theta}\partial_{\phi}^{2}+\cot\theta\partial_{\theta}-1\right)h^{J}=0.
\end{equation}
Therefore, we have to find \textit{everywhere} finite solutions of the following
equation 
\begin{equation}
\left({\vec{\nabla}}_{S^{2}}^{2}+k\right)f\left(\theta,\phi\right)=g\left(\theta,\phi\right),
\end{equation}
where ${\vec{\nabla}}_{S^{2}}^{2}$ is the Laplacian on $S^{2}$ given
as 
\begin{equation}
{\vec{\nabla}}_{S^{2}}^{2}:=\partial_{\theta}^{2}+\cot\theta\partial_{\theta}+\frac{1}{\sin^{2}\theta}\partial_{\phi}^{2}.
\end{equation}
One can employ the Green's function technique to solve this problem.
For the Helmholtz operator on the sphere, the Green function $G(\hat{x},\hat{x}')$
is defined as 
\begin{equation}
\left({\vec{\nabla}}_{S^{2}}^{2}+\lambda(\lambda+1)\right)G(\hat{x},\hat{x}')=\delta^{(2)}(\hat{x}-\hat{x}'),
\end{equation}
which can be found as an infinite series expansion (for example, see
\cite{Green}) 
\begin{equation}
G(\hat{x},\hat{x}')=\frac{1}{4\sin\pi\lambda}\sum_{n=0}^{\infty}\frac{1}{\left(n!\right)^{2}}\frac{\Gamma(n-\lambda)}{\Gamma(-\lambda)}\frac{\Gamma(n+\lambda+1)}{\Gamma(\lambda+1)}\left(\frac{1+\hat{x}\cdot\hat{x'}}{2}\right)^{n},
\end{equation}
where $\hat{x}=\sin\theta\cos\phi\hat{i}+\sin\theta\sin\phi\hat{j}+\cos\theta\hat{k}$
and $\hat{x}'$ is a similar expression with some other $\theta$
and $\phi$. Employing this Green's function with $\lambda=\frac{-1+i\sqrt{3}}{2}$,
one finds the first non-trivial correction to the location of the
Apparent Horizon as 
\begin{equation}
h^{p}=-\frac{1}{16}\left(c_{1}\sin\theta\cos\phi+c_{2}\cos\theta\right),
\end{equation}
and $h^{J}=0$. Therefore the apparent horizon is perturbed from the
zeroth order expansion to 
\begin{equation}
r=h(\theta,\phi)=a-\frac{p}{16}\Big(\sin\theta_{0}\sin\theta\cos\phi+\cos\theta_{0}\cos\theta\Big),
\end{equation}
where, recall that, $\theta_{0}$ is the angle between the linear
momentum and the spin vectors.  So the magnitude of the spin vector does is irrelevant at this order for the location of the Apparent Horizon, but the angle it makes with the momentum vector is relevant. In Figure 3, we plotted and example of how the shape of the horizon looks like. There is a dimple on the sphere whose size depends on the ratio $p/a$ which we took to be large to see the dimple.
In the limit $\theta_{0}=0$, $h$
reduces to the form given in \cite{Dennison-Baumgarte}. 

\begin{figure}
\centering \includegraphics[width=0.4\linewidth]{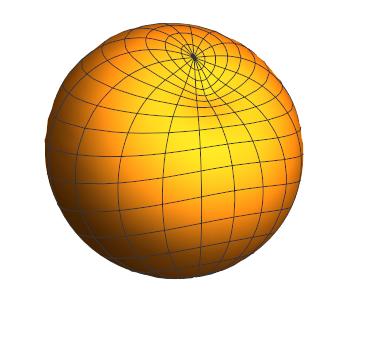}\caption{ Shape of the Apparent Horizon when the angle between $\vec{p}$ and
$\vec{J}$ is 45 degrees; and $p/a=8\sqrt{2}$ which is outside the
validity of the approximation we have worked with. }
\label{fig:AH1} 
\end{figure}
Let us now evaluate the area of the Apparent Horizon from (\ref{alan3})
which at the end yields 
\begin{equation}
A_{\text{AH}}=64\pi a^{2}+4\pi p^{2}+\frac{11\pi J^{2}}{5a^{2}}.
\end{equation}
Thus the irreducible mass $M_{\text{irr}}$ turns out to be 
\begin{equation}
M_{\text{irr}}=2a+\frac{p^{2}}{16a}+\frac{11J^{2}}{320a^{3}},
\end{equation}
so comparing with the energy, $E_{ADM}$, we have 
\begin{equation}
E_{\text{ADM}}=M_{\text{irr}}+\frac{p^{2}}{2M_{\text{irr}}}+\frac{J^{2}}{8M_{\text{irr}}^{3}},
\end{equation}
which matches the result (\ref{chris2}) of Christodoulou at this
order.

\section{Conclusions}

We have presented a step-by-step construction of the Apparent Horizon
equation which is of extreme importance in black hole physics; and
described in detail how it correctly yields the expected results,
such as the irreducible mass, for a slowly moving and spinning black
hole. For stationary black holes the event Horizon and the Apparent
Horizon coincide. This exposition is of a pedagogical nature with details given in the Appendix including the derivation of the null Raychaudhuri equation which we have not used in the text, but added for more insight for the expansion of a null geodesic.
We have skipped some interesting issues such as: numerically solving
the case with no symmetry; multi black hole initial data; the proof
that when the dominant energy condition is satisfied, the topology
of the Apparent Horizon is that of $S^{2}$. 
For other nice expositions regarding horizons and related concepts see \cite{Eric, Booth, Curiel}.

\begin{acknowledgments}
The work of E.A. is partially supported by the TUBITAK Grant No. 120F253. This work was written 
to celebrate the 70th birthday of Tekin Dereli who has been extremely influential in gravity research in Turkey. 
\end{acknowledgments}

\section{Appendix}

\subsection{An equivalent definition of the expansion $\Theta_{\left(l\right)}$}

Here we give a proof of the second equality in (\ref{proof1}): we
have 
\begin{equation}
\Theta_{\left(\ell\right)}=q^{\mu\nu}\nabla_{\mu}\ell_{\nu}=\frac{1}{2}q^{\mu\nu}{\mathcal{L}}_{\ell}q_{\mu\nu},
\end{equation}
where the first equality is identical to the definition of the $\Theta_{\left(\ell\right)}$.
Starting from $q^{\mu\nu}{\mathcal{L}}_{\ell}q_{\mu\nu}$, one can
easily arrive at the expansion $\Theta_{\left(\ell\right)}$. The construction
is as follows: 
\begin{equation}
q^{\mu\nu}{\mathcal{L}}_{\ell}q_{\mu\nu}=q^{\mu\nu}\left(\ell^{\sigma}\nabla_{\sigma}q_{\mu\nu}+q_{\sigma\nu}\nabla_{\mu}\ell^{\sigma}+q_{\sigma\mu}\nabla_{\nu}\ell^{\sigma}\right). \label{eq}
\end{equation}
The first term on the right hand side automatically vanishes. To be
able to see this explicitly, we express the metric $q_{\mu\nu}$ in
terms of the spacetime metric $g_{\mu\nu}$
\begin{equation}
q^{\mu\nu}\ell^{\sigma}\nabla_{\sigma}q_{\mu\nu}=q^{\mu\nu}\ell^{\sigma}\nabla_{\sigma}\left(g_{\mu\nu}+k_{\mu}\ell_{\nu}+k_{\nu}\ell_{\mu}\right).
\end{equation}
Since $\nabla_{\sigma}g_{\mu\nu}=0$, the non vanishing terms are
\begin{equation}
q^{\mu\nu}\ell^{\sigma}\nabla_{\sigma}q_{\mu\nu}=q^{\mu\nu}\ell^{\sigma}\left(k_{\mu}\nabla_{\sigma}l_{\nu}+\ell_{\nu}\nabla_{\sigma}k_{\mu}+k_{\nu}\nabla_{\sigma}\ell_{\mu}+\ell_{\mu}\nabla_{\sigma}k_{\nu}\right),
\end{equation}
where $q^{\mu\nu}k_{\mu}=0=q^{\mu\nu}\ell_{\mu}$, and so one gets 
\begin{equation}
q^{\mu\nu}l^{\sigma}\nabla_{\sigma}q_{\mu\nu}=0.
\end{equation}
Now let us evaluate the second and third terms in (\ref{eq}) (which contribute equally). We can write 
\begin{equation}
q^{\mu\nu}q_{\sigma\nu}\nabla_{\mu}\ell^{\sigma}=q^{\mu\nu}\left(g_{\sigma\nu}+k_{\sigma}\ell_{\nu}+k_{\nu}\ell_{\sigma}\right)\nabla_{\mu}\ell^{\sigma}.
\end{equation}
Using $q^{\mu\nu}k_{\mu}=0=q^{\mu\nu}\ell_{\mu}$ again, the last expression
reduces to the following 
\begin{equation}
q^{\mu\nu}q_{\sigma\nu}\nabla_{\mu}\ell^{\sigma}=q^{\mu\nu}g_{\sigma\nu}\nabla_{\mu}\ell^{\sigma}=q^{\mu\nu}\nabla_{\mu}\ell_{\nu}.
\end{equation}
Then (\ref{eq}) becomes 
\begin{equation}
q^{\mu\nu}{\mathcal{L}}_{\ell}q_{\mu\nu}=2q^{\mu\nu}\nabla_{\mu}\ell_{\nu},
\end{equation}
and one ends up with 
\begin{equation}
\frac{1}{2}q^{\mu\nu}{\mathcal{L}}_{\ell}q_{\mu\nu}=q^{\mu\nu}\nabla_{\mu}\ell_{\nu}=\Theta_{\left(\ell\right)},
\end{equation}
which is the expression we wanted to prove.

\subsection{Derivative of the area along the null vector field $\ell^{\mu}$}

To gain a better physical insight to the expansion of the null geodesic congruence, let us show that
when one takes the derivative of the area of the cross section along
the null vector field $\ell^{\mu}$, the expansion $\Theta_{\left(\ell\right)}$
can be directly obtained as the integrand. On the surface ${\mathcal{S}}$,
let us start with the area formula 
\begin{equation}
A=\intop dS\sqrt{q},
\end{equation}
of which the derivative along $\ell$ yields 
\begin{equation}
\ell^{\mu}\partial_{\mu}A=\intop dS\thinspace\ell^{\mu}\partial_{\mu}\sqrt{q},=\frac{1}{2}\intop dS\thinspace\sqrt{q}\thinspace\ell^{\mu}q^{ab}\partial_{\mu}q_{ab}.\label{derivative of area}
\end{equation}
Equivalently one can express the result in terms of the Lie derivative
along the vector field $\ell^{\mu}$ using 
\begin{equation}
q^{ab}{\mathcal{L}}_{\ell}q_{ab}=q^{ab}\ell^{\mu}\partial_{\mu}q_{ab}+2q^{ab}q_{\mu b}\partial_{a}\ell^{\mu}.\label{eqqqq}
\end{equation}
Since the null vectors $\ell^{\mu}$ and $k^{\mu}$ are the elements
of the compliment of the subspace ${\mathcal{S}}$, one has $k^{a}=0=\ell^{a}$.
By definition (\ref{metriconS}) we obtain 
\begin{equation}
q_{\mu}^{a}=\delta_{\mu}^{a}+k_{\mu}\ell^{a}+k^{a}\ell_{\mu}=\delta_{\mu}^{a},
\end{equation}
and similarly 
\begin{equation}
q^{a\mu}=g^{a\mu}+k^{\mu}\ell^{a}+k^{a}\ell^{\mu}=g^{a\mu}.
\end{equation}
So that $q^{ab}q_{\mu b}=\delta_{\mu}^{a}$; and the last term in
(\ref{eqqqq}) becomes 
\begin{equation}
q^{ab}q_{\mu b}\partial_{a}\ell^{\mu}=\delta_{\mu}^{a}\partial_{a}\ell^{\mu}=\partial_{a}\ell^{a}=0.
\end{equation}
Then $q^{ab}{\mathcal{L}}_{\ell}q_{ab}$ reduces to 
\begin{equation}
q^{ab}{\mathcal{L}}_{\ell}q_{ab}=q^{ab}\ell^{\mu}\partial_{\mu}q_{ab},
\end{equation}
which can be related to $\ell^{\mu}\partial_{\mu}\sqrt{q}$ via 
\begin{equation}
\ell^{\mu}\partial_{\mu}\sqrt{q}=\frac{1}{2}\sqrt{q}\thinspace q^{ab}{\mathcal{L}}_{\ell}q_{ab}.
\end{equation}
Now we can rewrite (\ref{derivative of area}) in terms of Lie derivative
\begin{equation}
\ell^{\mu}\partial_{\mu}A=\intop dS\thinspace\sqrt{q}\thinspace\frac{1}{2}\thinspace q^{ab}{\mathcal{L}}_{\ell}q_{ab}.\label{area1}
\end{equation}
In order to show the appearance of the expansion $\Theta_{\left(\ell\right)}$
explicitly, we should use the spacetime coordinates, recall that we
have $\Theta_{\left(l\right)}=q^{\mu\nu}{\mathcal{L}}_{\ell}q_{\mu\nu}/2$,
instead of the coordinates on the co-dimension two surface ${\mathcal{S}}$.
It is straightforward to write 
\begin{equation}
q_{\mu}^{a}q_{v}^{b}{\mathcal{L}}_{\ell}q_{ab}=\delta_{\mu}^{a}\delta_{v}^{b}{\mathcal{L}}_{\ell}q_{ab}={\mathcal{L}}_{\ell}q_{\mu\nu}.
\end{equation}
Multiplying this with $q_{\sigma}^{\mu}q_{\rho}^{\nu}$ one obtains
\begin{equation}
q_{\sigma}^{\mu}q_{\rho}^{\nu}q_{\mu}^{a}q_{v}^{b}{\mathcal{L}}_{\ell}q_{ab}=q_{\sigma}^{\mu}q_{\rho}^{\nu}{\mathcal{L}}_{\ell}q_{\mu\nu},
\end{equation}
which in terms of Kronecker delta functions reads 
\begin{equation}
q_{\sigma}^{\mu}q_{\rho}^{\nu}\delta_{\mu}^{a}\delta_{v}^{b}{\mathcal{L}}_{\ell}q_{ab}=q_{\sigma}^{\mu}q_{\rho}^{\nu}{\mathcal{L}}_{\ell}q_{\mu\nu},
\end{equation}
and yields 
\begin{equation}
q_{\sigma}^{a}q_{\rho}^{b}{\mathcal{L}}_{\ell}q_{ab}=q_{\sigma}^{\mu}q_{\rho}^{\nu}{\mathcal{L}}_{\ell}q_{\mu\nu}.
\end{equation}
We multiply the last identity with $q^{\sigma\rho}$. Then we find
the identity 
\begin{equation}
q^{\sigma\rho}\delta_{\sigma}^{a}\delta_{\rho}^{b}{\mathcal{L}}_{\ell}q_{ab}=q^{\sigma\rho}q_{\sigma}^{\mu}q_{\rho}^{\nu}{\mathcal{L}}_{\ell}q_{\mu\nu},
\end{equation}
and so one arrives at 
\begin{equation}
q^{ab}{\mathcal{L}}_{\ell}q_{ab}=q^{\sigma\rho}q_{\sigma}^{\mu}q_{\rho}^{\nu}{\mathcal{L}}_{\ell}q_{\mu\nu},
\end{equation}
where $q^{\sigma\rho}q_{\sigma}^{\mu}q_{\rho}^{\nu}=q^{\mu\nu}$.
Finally we end up with 
\begin{equation}
q^{ab}{\mathcal{L}}_{\ell}q_{ab}=q^{\mu\nu}{\mathcal{L}}_{\ell}q_{\mu\nu}.
\end{equation}
This proves that the expansion $\Theta_{\left(\ell\right)}$ appears
in the change of the area along the vector field $\ell^{\mu}$. The
final expression is therefore 
\begin{equation}
\ell^{\mu}\partial_{\mu}A=\intop dS\thinspace\sqrt{q}\thinspace\frac{1}{2}\thinspace q^{\mu\nu}{\mathcal{L}}_{\ell}q_{\mu\nu}=\intop dS\thinspace\sqrt{q}\thinspace\frac{1}{2}\thinspace q^{ab}{\mathcal{L}}_{\ell}q_{ab}=\intop dS\thinspace\sqrt{q}\thinspace\Theta_{\left(l\right)}.\label{area1-1}
\end{equation}
So setting $\Theta_{\left(\ell\right)}=0$ to define the Apparent Horizon
boils down to setting $\ell^{\mu}\partial_{\mu}A=0$.

\subsection{Null Raychaudhuri equation}

The form of (\ref{proof1}) already suggests that one can define a
tensor whose trace is the expansion. Here we explore this tensor and
obtain an expression for the change of the null expansion along the
null direction $\ell$ as well as the null Raychaudhuri equation \cite{Eric0}.
So let us introduce the \textit{deformation tensor} $\Theta_{\mu\nu}$
as 
\begin{equation}
\Theta_{\mu\nu}:=\frac{1}{2}q_{\mu}^{\sigma}q_{\nu}^{\rho}{\mathcal{L}}_{\ell}q_{\sigma\rho},\label{aden1}
\end{equation}
such that 
\begin{equation}
\Theta_{\left(\ell\right)}=g^{\mu\nu}\Theta_{\mu\nu}=q^{\mu\nu}\nabla_{\mu}\ell_{\nu}.\label{adenk2}
\end{equation}
Carrying out the Lie-derivative in (\ref{aden1}), one has 
\begin{equation}
\Theta_{\mu\nu}=\nabla_{\mu}\ell_{\nu}-\omega_{\mu}\ell_{\nu}+\ell_{\mu}k^{\sigma}\nabla_{\sigma}\ell_{\nu},\label{deformation rate}
\end{equation}
with 
\begin{equation}
\omega_{\mu}:=-k^{\sigma}\nabla_{\mu}\ell_{\sigma}-k^{\sigma}k^{\rho}\ell_{\mu}\nabla_{\sigma}\ell_{\rho}\label{rotation}
\end{equation}
which is called the rotation one form.

In what follows, we will make use of the Ricci identity 
\begin{equation}
\nabla_{\mu}\nabla_{\nu}\ell^{\mu}-\nabla_{\nu}\nabla_{\mu}\ell^{\mu}=R_{\nu\lambda}\ell^{\lambda}.\label{Ricciidentity}
\end{equation}
From (\ref{adenk2}), one has 
\begin{equation}
\Theta_{\left(\ell\right)}=\nabla_{\mu}\ell^{\mu}+k^{\nu}\ell^{\mu}\nabla_{\mu}\ell_{\nu}.
\end{equation}
Here we assume that $\ell$ is a geodesic null vector but not necessarily
affinely parameterized so that 
\begin{equation}
\ell^{\mu}\nabla_{\mu}\ell_{\nu}=\kappa\ell_{\nu},
\end{equation}
where $\kappa$ is a function on spacetime. Using $k^{\nu}\ell_{\nu}=-1$,
one has 
\begin{equation}
\Theta_{\left(l\right)}=\nabla_{\mu}\ell^{\mu}-\kappa.
\end{equation}
So we have the following two equations: 
\begin{equation}
\nabla_{\mu}\ell^{\mu}=\Theta_{\left(\ell\right)}+\kappa,\hskip1cm\nabla_{\mu}\ell_{\nu}=\Theta_{\mu\nu}+\omega_{\mu}\ell_{\nu}-\ell_{\mu}k^{\sigma}\nabla_{\sigma}\ell_{\nu}.\label{two}
\end{equation}
Substituting these in (\ref{Ricciidentity}), one has 
\begin{equation}
\nabla_{\mu}\left(\Theta_{\nu}\thinspace^{\mu}+\omega_{\nu}\ell^{\mu}-\ell_{\nu}k^{\sigma}\nabla_{\sigma}\ell^{\mu}\right)-\nabla_{\nu}(\Theta_{\left(\ell\right)}+\kappa)=R_{\nu\lambda}\ell^{\lambda},
\end{equation}
which more explicitly becomes 
\begin{equation}
\nabla_{\mu}\Theta_{\nu}\thinspace^{\mu}+\ell^{\mu}\nabla_{\mu}\omega_{\nu}+\omega_{\nu}\nabla_{\mu}\ell^{\mu}-k^{\sigma}\nabla_{\sigma}\ell^{\mu}\nabla_{\mu}\ell_{\nu}-\ell_{\nu}\nabla_{\mu}(k^{\sigma}\nabla_{\sigma}\ell^{\mu})-\nabla_{\nu}(\Theta_{\left(\ell\right)}+\kappa)=R_{\nu\lambda}\ell^{\lambda}.
\end{equation}
We use the expressions (\ref{two}) one more time and reexpress the
third and the fourth terms to obtain 
\begin{eqnarray}
 &  & \nabla_{\mu}\Theta_{\nu}\thinspace^{\mu}+\ell^{\mu}\nabla_{\mu}\omega_{\nu}+\omega_{\nu}(\Theta_{\left(\ell\right)}+\kappa)-k^{\sigma}\nabla_{\sigma}\ell^{\mu}\left(\Theta_{\mu\nu}+\omega_{\mu}\ell_{\nu}-\ell_{\mu}k^{\gamma}\nabla_{\gamma}\ell_{\nu}\right)\nonumber \\
 &  & \hskip3cm-\ell_{\nu}\nabla_{\mu}(k^{\sigma}\nabla_{\sigma}\ell^{\mu})-\nabla_{\nu}(\Theta_{\left(l\right)}+\kappa)=R_{\nu\lambda}\ell^{\lambda}.
\end{eqnarray}
Since $\ell_{\mu}\nabla_{\sigma}\ell^{\mu}=0$, the last term on the
first line automatically vanishes. Contracting the final expression
with $\ell^{\nu}$ and using the fact that it is a null vector, one
arrives at 
\begin{equation}
\ell^{\nu}\nabla_{\mu}\Theta_{\nu}\thinspace^{\mu}+\ell^{\nu}\ell^{\mu}\nabla_{\mu}\omega_{\nu}+\ell^{\nu}\omega_{\nu}(\Theta_{\left(\ell\right)}+\kappa)-\ell^{\nu}\Theta_{\mu\nu}k^{\sigma}\nabla_{\sigma}\ell^{\mu}-\ell^{\nu}\nabla_{\nu}(\Theta_{\left(\ell\right)}+\kappa)=R_{\nu\lambda}\ell^{\lambda}\ell^{\nu}.\label{three}
\end{equation}
It is easy to show that the contraction of the null vector $\ell^{\nu}$
and the deformation tensor identically vanishes, $\ell^{\nu}\Theta_{\nu}\thinspace^{\mu}=0$.
Then one has 
\begin{equation}
\ell^{\nu}\nabla_{\mu}\Theta_{\nu}\thinspace^{\mu}=-\Theta_{\mu\nu}\Theta^{\mu\nu},
\end{equation}
and also 
\begin{equation}
\ell^{\nu}\ell^{\mu}\nabla_{\mu}\omega_{\nu}=\ell^{\mu}\nabla_{\mu}\kappa-\kappa^{2}.
\end{equation}
Inserting these expressions in (\ref{three}) we get 
\begin{equation}
-\Theta_{\mu\nu}\Theta^{\mu\nu}+\kappa\Theta_{\left(l\right)}-\ell^{\nu}\nabla_{\nu}\Theta_{\left(l\right)}=R_{\nu\lambda}\ell^{\lambda}\ell^{\nu}.\label{four}
\end{equation}
The first term $\Theta_{\mu\nu}\Theta^{\mu\nu}$ can be written in
terms of the trace free shear tensor $\sigma_{\mu\nu}$ 
\begin{equation}
\sigma_{\mu\nu}:=\Theta_{\mu\nu}-\frac{1}{n-2}q_{\mu\nu}\Theta_{\left(\ell\right)}
\end{equation}
as follows 
\begin{equation}
\Theta_{\mu\nu}\Theta^{\mu\nu}=\mathbf{\sigma_{\mu\nu}\sigma^{\mu\nu}+}\frac{1}{n-2}\Theta_{\left(\ell\right)}^{2},
\end{equation}
where $\sigma_{\mu\nu}\sigma^{\mu\nu}=\sigma_{ab}\sigma^{ab}$. Then
(\ref{four}) becomes 
\begin{equation}
\ell^{\mu}\nabla_{\mu}\Theta_{\left(\ell\right)}=\nabla_{\ell}\Theta_{(\ell)}=\kappa\Theta_{\left(l\right)}-R_{\mu\nu}\ell^{\mu}\ell^{\nu}-\sigma_{ab}\sigma^{ab}-\frac{1}{n-2}\Theta_{\left(\ell\right)}^{2}.
\end{equation}
The null vector field $\ell^{\mu}$ is oriented in the future direction.
Therefore the last equation, known as the null Raychaudhuri
equation,  is an evolution equation for the expansion
$\Theta_{\left(\ell\right)}$.

\end{document}